\documentclass[preprint,aps]{revtex4}
\usepackage{amssymb}

\usepackage{graphicx}

\begin{document}
\title{Nonlocal impedances and the Casimir entropy at low temperatures}
\author{V. B. Svetovoy}
\email[]{V.B.Svetovoy@el.utwente.nl}
\thanks{On leave from Yaroslavl University, Yaroslavl, Russia}
\affiliation{MESA+ Research Institute, University of Twente, P.O.
217, 7500 AE Enschede, The Netherlands}
\author{R. Esquivel}
\email[]{raul@fisica.unam.mx} \affiliation{Instituto de Fisica,
Universidad Nacional Aut\'{o}noma de M\'{e}xico, Apartado Postal
20-364, DF 01000 M\'{e}xico, Mexico}

\date{\today}

\begin{abstract}
The problem with the temperature dependence of the Casimir force is
investigated. Specifically, the entropy behavior in the low
temperature limit, which caused debates in the literature, is
analyzed. It is stressed that the behavior of the relaxation
frequency in the $T\rightarrow0$ limit does not play a physical role
since the anomalous skin effect dominates in this range. In contrast
with the previous works, where the approximate Leontovich impedance
was used for analysis of nonlocal effects, we give description of
the problem in terms of exact nonlocal impedances. It is found that
the Casimir entropy is going to zero at $T\rightarrow0$ only in the
case when $s$ polarization does not contribute to the classical part
of the Casimir force. However, the entropy approaching zero from the
negative side that, in our opinion, cannot be considered as
thermodynamically satisfactory. The resolution of the negative
entropy problem proposed in the literature is analyzed and it is
shown that it cannot be considered as complete. The crisis with the
thermal Casimir effect is stressed.
\end{abstract}

\pacs{}
\maketitle

\section{Introduction\label{sec1}}

An attractive force between uncharged metallic plates, predicted in
1948 by Casimir \cite{Cas48}, is one of the most striking
macroscopic manifestations of quantum vacuum. Recently this force
became a subject of systematic experimental investigation
\cite{Lam97,Moh98,Ede00,Cha01,Bre02,Dec03,Dec05,Ian04}. The force
between ideal metals at zero temperature \cite{Cas48},

\begin{equation}
F=-\frac{\pi ^{2}\hbar c}{240a^{4}},  \label{Cas}
\end{equation}

\noindent depends only on the separation $a$ and fundamental
constants. In reality the force is measured at finite temperature
between deposited metallic films, which have finite conductivity and
roughness. Correction to Eq. (\ref{Cas}) due to finite conductivity
can be as large as 50\% for small separations $a\sim 100\ nm$.
Contribution of the finite temperature to this correction is not
large but caused a lot of controversy in the literature (see Ref.
\cite{Mil04} for a recent review). The essence of the problem lies
in the classic contribution to the Casimir force, which dominates at
large distances between plates or at high temperature. Calculations
made for ideal metals at finite temperature \cite {Meh67,Bro69}
showed that $s$- and $p$-polarized modes of electromagnetic field
gave equal contributions to the force. At the same time the Lifshitz
theory of fluctuating fields \cite{Lif56,LP9} predicted zero
contribution for $s$-polarization. For the first time the problem
was recognized many years ago. For reconciliation of the results
Schwinger, DeRaad, and Milton (SDM) \cite{Sch78} proposed a special
prescription to be used with the Lifshitz formula: one has to take
first the limit $\varepsilon \rightarrow \infty $ for the metal
permittivity and only then allow the frequency $\omega $ go to zero.
Modern calculations concerned with nonideal metals were  confronted with
the problem again.

Different approaches to resolve the problem have been proposed in
the literature, which resulted in different temperature corrections
to the Casimir force. Bostr\"{o}m and Sernelius \cite{Bos00} used
the Lifshitz formula with the Drude dielectric function and found
that $s$-polarization did not contribute in the classical limit
($n=0$ term in the Lifshitz formula) independently on the Drude
parameters. In this approach there is no continuous transition to
the ideal metal case and the predicted temperature correction is in
contradiction with the Lamoreaux experiment \cite{Lam97}. However,
physically this approach is well motivated since the Drude
dielectric function is working especially well at low frequencies.
Bordag et al. \cite{Bor00} used the plasma model dielectric
function, for which $\varepsilon $ at low frequencies increases
faster ( $\omega ^{-2}$) than for the Drude model ($\omega ^{-1}$).
They found that $s$-polarization gives finite contribution in the
classical limit, which coincides with the ideal metal result when
the plasma frequency $ \omega _{p}$\ is going to infinity. The
temperature correction in this approach is very close to that for
the ideal metal and negligible at small separations between plates.
A weak point of this approach is that no known material behaves at
low frequency according to the plasma model. Svetovoy and Lokhanin
\cite{Sve00} proposed to use SDM prescription for the $n=0$ term in
the Lifshitz formula for real metals also. Later it was shown \cite
{Sve01} that this prescription follows from very general dimensional
analysis of the classical contribution to the force if one demands
continuous transition to the ideal metal case. The temperature
correction happened to be small but observable at small separations
between bodies.

A new round of discussion has started when a thermodynamical problem
connected with the Casimir free energy has been revealed
\cite{Bez02a}. The idea was to use the Nernst heat theorem as a
guiding principle to choose between different approaches to the
temperature correction. According to this theorem the entropy has to
go to zero in the limit of zero temperature. It was noted
\cite{Bez02a} that the Drude relaxation frequency $\omega _{\tau }$\
vanishes with $T$ \ and, therefore, the plasma dielectric function
is realized at $T\rightarrow 0$. In this case the leading term in
the temperature correction is $\sim T^{3}$ \cite{Bor00} and the
entropy is safely going to zero as $S\sim T^{2}$. Two other
approaches predict the leading term in the correction $\sim T$ and
finite entropy at $ T=0$, positive and negative for the approaches
\cite{Sve00} and \cite{Bos00}, respectively. However, the following
analysis revealed that the situation is not as simple. The anomalous
skin effect was shown to be important for the temperature correction
at low temperatures \cite{Sve03}. With the use of the Leontovich
impedance for the anomalous skin effect it was demonstrated that the
entropy is going to zero only if SDM prescription is used for the $
n=0$ term. On the other hand, it was noted \cite{Hoy03} and
expressed later more clearly \cite{Bos04} that any real material
contains a number of defects, which are responsible for the residual
resistance at $T=0$. Equivalently it means that $\omega _{\tau }$
becomes very small but finite at $T=0$. It was shown, that in this
case, the entropy disappears at sufficiently low temperature
\cite{Hoy03,Bos04}. Therefore, again we have a confusing
situation where each approach has its own reasoning.

We would like to emphasize that at low temperatures the anomalous
skin effect plays an important role and should be taken into account
in any reasonable calculations. Because $\omega _{\tau }$ decreases
fast with the temperature, at sufficiently low temperature
inevitably the mean free path $ l=v_{F}/\omega _{\tau }(T)$ for
electrons becomes much larger then the field penetration depth
$\delta $. When this happens the relaxation frequency does not play a
physical role any more. Instead of $\omega _{\tau }$ the physical
significance gets the other frequency, $\Omega =\left(
v_{F}/c\right) \omega _{p}$, which is often used as a characteristic
frequency of anomalous skin effect. For this reason the question,
does $\omega _{\tau }$ go to zero or have some residual value at
$T\rightarrow 0$, becomes unimportant. This ideology was developed
in Ref. \cite{Sve03} in context of the thermal correction to the
Casimir force. For the description of the anomalous skin effect the
Leontovich impedance was used there.

The approximate Leontovich impedance was used for calculations
\cite{Bez02b,Sve03,Gey03} (see additional discussion in Refs.
\cite{Bez04,Sve04,Gey04}). The approach similar to the Leontovich
impedance was developed also in Refs. \cite{Tor04,Lam05}. This
impedance describes well the propagating electromagnetic field, but
it was not clear why in the local limit it gives the result
different from the dielectric function approach \cite{Sve03}. In
Refs. \cite{Moc02,Esq03,Esq05} it was demonstrated that the use of the
exact impedances is in agreement with the dielectric function
approach. It became clear that the point of contradiction is the
transverse momentum, which is neglected in the Leontovich impedance
\cite{Bre03,Bre04,Bez04}. In our paper \cite{Esq04} a general
approach to the nonlocal impedances was developed for applications
in the Casimir force calculations. It was shown that for real metals
both contributions in the force from propagating and evanescent
fields are important. The propagating fields can be described well
by the Leontovich impedance, but the same is not true for the
evanescent fields. The latter ones should be described by more
general impedances, for which dependence on the transverse momentum
cannot be neglected. Explicit expressions for these impedances were
presented in Ref. \cite{Esq04}.   

It is important to notice that the relevance of spatial dispersion effects depends on the separation between the slabs, being more important at short separations.   For two Au slabs, at a separation of the order of the plasma wavelength of Au($130-140 nm$), the difference between the local and nonlocal calculation is $0.2\%$, and can be significant when experimental errors of the order of  $.5\%$ are claimed \cite{Dec05}. For the hydrodynamic model ref. \cite{Esq03}and ref. \cite{Hein75} give the same results, using a dielectric function valid in a wide range of frequencies not only at the infrared as stated in \cite{Dec05}.

Inadequacy of the Leontovich impedance forced us to reconsider the
result of Ref. \cite{Sve03} for the entropy behavior in the low
temperature limit. In this paper we calculate analytically the 
temperature correction to the Casimir free energy using the general
approach to the nonlocal impedances \cite{Esq04}.

The paper is organized as follows. In Sec. \ref{sec2} we separate
the temperature dependent part of the free energy and transform it
to the form convenient for calculations. In Sec. \ref{sec3} the
nonlocal impedances at low temperature are discussed. In Sec.
\ref{sec4} we give analytic expressions for for the free energy in
two limit cases. The entropy behavior at $T\rightarrow0$ and
discussion around are given in Sec. \ref{sec5}. In the last section
we present our conclusions.

\section{Temperature dependent part of the free energy\label{sec2}}

The Casimir force at nonzero temperature between plates made of real
materials is given by the Lifshitz formula \cite{LP9}. For the free
energy $ {\cal F}(a,T)$ this formula can be presented in the
following form

\begin{equation}
{\cal F}(a,T)=\frac{kT}{8\pi a^{2}}\sum_{n=0}^{\infty }{}^{\prime
}\int\limits_{\xi _{n}}^{\infty }dy\,y\left\{ \ln \left[ 1-r_{s}^{2}\left(
\xi _{n},y\right) e^{-y}\right] +\left( r_{s}\rightarrow r_{p}\right)
\right\} ,  \label{fren}
\end{equation}

\noindent where $\xi_{n}$ are the dimensionless Matsubara
frequencies defined with respect to the characteristic frequency
$\omega _{a}$

\begin{equation}
\xi _{n}=\frac{\zeta _{n}}{\omega _{a}},\quad \zeta _{n}=\frac{2\pi
kT}{ \hbar }n,\quad \omega _{a}=\frac{c}{2a}.  \label{ksidef}
\end{equation}

\noindent In Eq. (\ref{fren}) $r_{s}$ and $r_{p}$ are the reflection
coefficients for $s$ and $p$-polarizations, respectively. The integration
variable $y$ is defined via the physical values as

\begin{equation}
y=2a\sqrt{\zeta _{n}^{2}/c^{2}+q^{2}},  \label{ydef}
\end{equation}

\noindent where $q$ is the absolute value of the wave vector along the plate.

The problem with the thermal correction comes from the $n=0$ term in
Eq. (\ref{fren}), which will be denoted as ${\cal F}_{0}\left(
a,T\right) $. There is no agreement between different authors
\cite{Bos00,Bor00,Sve00} what is the reflection coefficient
$r_{s}\left( 0,y\right) $ in this term. The $n=0$ term describes the
classical contribution to the free energy, which dominates at large
separations or high temperatures. Without loss of generality it can
be parameterized as

\begin{equation}
{\cal F}_{0}\left( a,T\right) =-\alpha \frac{kT}{8\pi a^{2}}\zeta \left(
3\right) .  \label{zeroterm}
\end{equation}

\noindent Here $\alpha $ is a dimensionless function of material
parameters and separation $a$. This form of ${\cal F}_{0}$ follows
from a simple dimensional analysis \cite{Sve01} in the classical
limit (not considering the Plank constant $\hbar $). Different approaches to the
temperature correction problem give different values of $\alpha $.
This value will be kept arbitrary in the calculations and will be
specified only for the discussion of the final result.

We are interested in the temperature dependent part of the free energy,
which is responsible for the entropy. To separate the temperature
independent part, let us rewrite the free energy in the following form

\begin{equation}
{\cal F}(a,T)={\cal F}_{0}\left( a,T\right) +\frac{\hbar c}{16\pi a^{3}}%
\frac{\tau }{2\pi }\sum\limits_{n=1}^{\infty }\left[ G_{s}\left( n\tau
\right) +G_{p}\left( n\tau \right) \right] ,  \label{freeG}
\end{equation}

\noindent where the functions $G_{i}\left( n\tau \right) $ ($i=s,p$) are
defined as

\begin{equation}
G_{i}\left( n\tau \right) =\int\limits_{n\tau }^{\infty }dy\,y\ln \left[
1-r_{i}^{2}\left( n\tau ,y\right) e^{-y}\right]  \label{Gdef}
\end{equation}

\noindent and $n\tau $ was introduced instead of $\xi _{n}$. The
parameter $ \tau $,

\begin{equation}
\tau =\frac{2\pi T}{T_{eff}},\ kT_{eff}=\frac{\hbar c}{2a}=\hbar \omega _{a},
\label{taudef}
\end{equation}

\noindent is a  dimensionless temperature. It is convenient to
rewrite the sum in Eq. (\ref{freeG}) using the Abel-Plana formula

\begin{equation}
\frac{\tau }{2\pi }\sum\limits_{n=1}^{\infty }G_{i}\left( n\tau
\right) = \frac{1}{2\pi }\int\limits_{0}^{\infty }G_{i}\left(
x\right) dx+\frac{\tau }{ 2\pi }\left[ \frac{1}{2}G_{i}\left( \tau
\right) -\int\limits_{0}^{1}G_{i}\left( \tau t\right) dt-2
\mathop{\rm Im} \int\limits_{0}^{\infty }\frac{G_{i}\left( \tau
+it\tau \right) }{e^{2\pi t}-1}dt\right] .  \label{APF}
\end{equation}

\noindent The first term on the right hand side does not depend on
temperature, but all the other terms describe the temperature correction.
The temperature dependent part of the free energy $\Delta {\cal F}\left(
a,T\right) ={\cal F}\left( a,T\right) -{\cal F}\left( a,0\right) $ can be
presented then in the following form:

\begin{equation}
\Delta {\cal F}\left( a,T\right) ={\cal F}_{0}\left( a,T\right)
+\frac{kT}{ 8\pi a^{2}}\left[ \frac{1}{2}G_{s}\left( \tau \right)
-\int\limits_{0}^{1}G_{s}\left( \tau t\right) dt-2 \mathop{\rm Im}
\int\limits_{0}^{\infty }\frac{G_{s}\left( \tau +it\tau \right)
}{e^{2\pi t}-1}dt+(s\rightarrow p)\right] .  \label{delF}
\end{equation}

It is important to see clearly which frequencies give the main
contribution to $\Delta {\cal F}\left( a,T\right) $. Indeed, the
most important contribution to the temperature independent part
comes from the Matsubara frequencies $\zeta _{n}\sim \omega _{a}$ or
$n\tau \sim 1$. The same is not true for $\Delta {\cal F}\left(
a,T\right) $. As one can see from Eq. (\ref {delF}) the important
values of the dimensionless frequency $\xi =\tau t$ are of the order
of 1 or $\zeta \sim \tau \omega _{a}$. We are analyzing the
temperature behavior in the low temperature range, where $\tau \ll
1$. Therefore, frequencies much smaller than the characteristic
frequency $ \omega _{a}$ give the main contribution to the
temperature dependent part of the the free energy.

\section{Nonlocal impedances at low frequencies\label{sec3}}

As was said in Introduction, at low temperatures the importance of the
anomalous skin effect significantly increases. Description of this
effect is given within the theory of nonlocal interaction between
the electromagnetic field and a metal. In this theory, the  reflectivity
of the metal is described by the surface impedances. The impedances
are connected with the nonlocal dielectric functions by the general
relations \cite{Kli68}

\begin{equation}
Z_{s}\left( \omega ,q\right) =\frac{i}{\pi }\frac{\omega }{c}%
\int\limits_{-\infty }^{\infty }\frac{dk_{z}}{\left( \omega
^{2}/c^{2}\right) \varepsilon _{t}-k^{2}},  \label{Zsreal}
\end{equation}

\begin{equation}
Z_{p}\left( \omega ,q\right) =\frac{i}{\pi }\frac{\omega }{c}%
\int\limits_{-\infty }^{\infty }\frac{dk_{z}}{k^{2}}\left[ \frac{q^{2}}{%
\left( \omega ^{2}/c^{2}\right) \varepsilon _{l}}+\frac{k_{z}^{2}}{\left(
\omega ^{2}/c^{2}\right) \varepsilon _{t}-k^{2}}\right] ,  \label{Zpreal}
\end{equation}

\noindent where $k=\sqrt{q^{2}+k_{z}^{2}}$ is the wave number,
$\varepsilon _{t}\left( k,\omega \right) $ and $\varepsilon
_{l}\left( k,\omega \right) $ are the nonlocal dielectric functions
describing the material response to transverse and longitudinal
electric fields, respectively. These equations are true
independently on the particular model used for obtaining $
\varepsilon _{t}$ and $\varepsilon _{l}$. The dielectric functions
can be found, for example, by solving the Boltzmann kinetic equation
\cite {Kli68}. The Boltzmann approximation is valid in the range
$\omega <\omega _{p}$ and $q<k_{F}$, where $k_{F}$ is the Fermi wave
number, and is appropriate for our problem.

The impedances Eq.(\ref{Zsreal}) and Eq.(\ref{Zpreal}) are analytic functions
in the upper half of the complex frequency plane and can be written
at imaginary frequencies $\omega =i\zeta $ using the analytic
continuation. Explicit form of the dielectric functions along the imaginary
axis \cite{Esq04} is

\begin{equation}
\varepsilon _{l}\left( \zeta ,v\right) =1+\frac{\omega _{p}^{2}f_{l}\left(
v\right) }{\zeta \left( \zeta +\omega _{\tau }\right) },\quad f_{l}\left(
v\right) =\frac{3}{v^{2}}\cdot \frac{v-\arctan v}{v+\left( \omega _{\tau
}/\zeta \right) \left( v-\arctan v\right) },  \label{El}
\end{equation}

\begin{equation}
\varepsilon _{t}\left( \zeta ,v\right) =1+\frac{\omega _{p}^{2}f_{t}\left(
v\right) }{\zeta \left( \zeta +\omega _{\tau }\right) },\quad f_{t}\left(
v\right) =\frac{3}{2v^{3}}\left[ -v+\left( 1+v^{2}\right) \arctan v\right] ,
\label{Et}
\end{equation}

\begin{equation}
v=v_{F}\frac{k}{\zeta +\omega _{\tau }},  \label{vdef}
\end{equation}

\noindent where $v_{F}$ is the Fermi velocity. The range of the anomalous
skin effect corresponds to large values of $v$. When the Casimir force is
calculated, $k$ is restricted by the condition $k\geqslant q\sim 1/2a$. On
the other hand, the denominator in Eq. (\ref{vdef}) is small and the
condition $v\gg 1$ will be fulfilled at sufficiently low temperature. In
this limit the dielectric functions behave as

\begin{equation}
\varepsilon _{l}\left( \zeta ,k\right) =1+3\left( \frac{\omega _{p}}{v_{F}k}%
\right) ^{2},  \label{Elanom}
\end{equation}

\begin{equation}
\varepsilon _{t}\left( \zeta ,k\right) =1+\frac{3\pi }{4}\frac{\omega
_{p}^{2}}{\zeta v_{F}k}.  \label{Etanom}
\end{equation}

\noindent One can immediately see that the relaxation frequency
falls out from the dielectric functions. The longitudinal function,
$\varepsilon _{l}$, does not depend on frequency at all, but $k$
dependence describes the Thomas-Fermi screening of the longitudinal
electric field. In the transverse function, $\varepsilon _{t}$, the term $v_{F}k$ plays a
role of the relaxation frequency. The surface
impedances corresponding to the functions (\ref{Elanom}), (\ref
{Etanom}) were found in Ref. \cite{Esq04}:

\begin{equation}
Z_{s}\left( \zeta ,q\right) =\frac{\zeta }{cq}F(b),  \label{Zs}
\end{equation}

\begin{equation}
Z_{p}\left( \zeta ,q\right) =\frac{q^{2}}{\sqrt{3}}\frac{cv_{F}}{\zeta
\omega _{p}}+\frac{\zeta }{cq}G(b),  \label{Zp}
\end{equation}

\noindent where the functions $F\left( b\right) $ and $G\left( b\right) $
are defined as

\begin{equation}
F(b)=\frac{2}{\pi }\int\limits_{0}^{\infty }d\chi \frac{\cosh
^{2}\chi }{ \cosh ^{3}\chi +b^{3}},\quad G\left( b\right)
=\frac{2}{\pi } \int\limits_{0}^{\infty }d\chi \frac{\sinh ^{2}\chi
}{\cosh ^{3}\chi +b^{3}}, \label{FGdef}
\end{equation}

\begin{equation}
b=\frac{1}{q}\left( \frac{3\pi }{4}\frac{\omega _{p}^{2}\zeta
}{c^{2}v_{F}} \right) ^{1/3}.  \label{bpar}
\end{equation}

\noindent The asymptotics for large and small values of $b$ are

\begin{equation}\label{asymsmall}
    F(b)=1+O(b^3),\ G(b)=\frac{1}{2}+O(b^3),\ b\ll1,
\end{equation}

\begin{equation}\label{asymlarge}
    F(b)=\frac{4}{3\sqrt{3}}\frac{1}{b}+O(b^{-3}),\
    G(b)=\frac{4}{3\sqrt{3}}\frac{1}{b}+O(b^{-3}),\ b\gg1.
\end{equation}

The Leontovich impedance for the strong anomalous skin effect
\cite{Abr88} is reproduced at finite frequency in the limit
$q\rightarrow0$ when $b\gg1$:

\begin{equation}
Z_{s}(0,\zeta )=Z_{p}\left( 0,\zeta \right) =Z\left( \zeta \right)
=\frac{4}{3\sqrt{3}}\left( \frac{4}{3\pi }\frac{v_{F}}{c}\frac{\zeta
^{2}}{\omega_{p}^2}\right) ^{1/3}. \label{Leont}
\end{equation}

\noindent However, the most important contribution to the Casimir
force give finite values of $q\sim1/2a$ and the limit $b\gg1$
inevitably will be broken at some sufficiently low frequency
(temperature). When $\zeta $ is so small that $b\ll 1$, the
impedance $Z_{s}$ approaches the local limit $Z_s(q,\zeta)=\zeta
/cq$, which does not depend on $\omega_{\tau}$. This is in contrast
with the Leontovich impedance, which behaves in the local limit as
$Z_s(\zeta)=\varepsilon(i\zeta)^{-1/2}\rightarrow\sqrt{\zeta
\omega_{\tau}/\omega_{p}^{2}}$. It clearly depends on the value of
$\omega_{\tau}$. Indeed, the reason for this is the non-dependence on $q$
 of the Leontovich approximation. The impedance $Z_p$
also behaves very differently from the Leontovich impedance in the
limit $b\ll1$, but in contrast with $Z_{s}$ it is significantly
nonlocal. This is because for $b\ll1$ the main contribution in $Z_p$
gives the first term in Eq. (\ref{Zp}) responsible the Thomas-Fermi
screening.

Let us discuss now the temperature range where Eqs. (\ref{Zs}) and
(\ref{Zp}) are true. The main condition is that the parameter $v$ in
Eq. (\ref{vdef}) has to be large. The minimal value of the wave
number is $k=q\sim1/2a$. The important frequencies contributing to
the temperature dependent part of the free energy Eq.(\ref{delF}) are
$\zeta\sim 2\pi kT/\hbar$. We assume that the relaxation frequency
$\omega_{\tau}$ decreases with temperature faster than linearly and
for this reason  it can be neglected in the denominator of Eq.
(\ref{vdef}). At very low temperatures this assumption can be
broken due to residual resistivity, but in this case $v$ will be
certainly large. Therefore, the value of $v$ will be much larger
than 1 if

\begin{equation}\label{Tcond1}
    kT\ll\frac{\hbar\omega_a}{2\pi}\frac{v_F}{c}.
\end{equation}

\noindent When this condition is met, the impedances Eq.(\ref{Zs}),
Eq.(\ref{Zp}) can be used independently on the value of $b$.
However, for large $b$, when the Leontovich impedance (\ref{Leont})
can be used, the condition on the temperature is relaxed. This is
because for $b\gg1$, the wave number $k=q\cosh\chi\sim qb\gg1/2a$.
In this limit the condition $v\gg1$ means

\begin{equation}\label{Tcond2}
    kT\ll\frac{\hbar\omega_p}{4}\sqrt{\frac{3}{\pi}}\frac{v_F}{c}
\end{equation}

\noindent This is the restriction on the temperature used in Ref.
\cite{Sve03}. Therefore, the Leontovich impedance can be used in the
temperature range $\hbar\omega_a(v_F/c)\ll 2\pi
kT\ll\sqrt{3/16\pi}\hbar\omega_p(v_F/c)$. When the temperature is
going down and obeys the condition $2\pi kT\ll\hbar\omega_a(v_F/c)$, the
$q$-dependence of the impedances becomes important and one has to
use Eqs. (\ref{Zs}), (\ref{Zp}), and Eq. (\ref{asymsmall}) instead of
Eq.(\ref{Leont}).

\section{Evaluation of the free energy\label{sec4}}
The temperature dependent part of the free energy is defined in Eq.
(\ref{delF}), where the functions $G_{s,p}$ are given by Eq.
(\ref{Gdef}). The reflection coefficients $r_{s,p}$ can be expressed
via the impedances as

\begin{equation}\label{refl}
    r_s=-\frac{Z_{s0}-Z_s}{Z_{s0}+Z_s}, \quad
    r_p=\frac{Z_{p0}-Z_p}{Z_{p0}+Z_p},
\end{equation}

\noindent where $Z_{s0}$ and $Z_{p0}$ are the "impedances" of the
plain wave defined as the ratios of the electric and magnetic fields
in the wave:

\begin{equation}\label{Z0}
    Z_{s0}=\frac{\zeta}{ck_0}, \quad Z_{p0}=\frac{ck_0}{\zeta},
    \quad k_0 = \sqrt{\zeta^2 /c^2 +q^2}.
\end{equation}

Let us first calculate the function $G_s(\tau)$. Using Eq.
(\ref{Zs}) for the impedance $Z_s$ the reflection coefficient can be
written in the following form

\begin{equation}\label{refls}
    r_s=-\frac{1-\frac{y}{\sqrt{y^2-\tau^2}} F(A/\sqrt{y^2-\tau^2})}
    {1+\frac{y}{\sqrt{y^2-\tau^2}} F(A/\sqrt{y^2-\tau^2})},
\end{equation}

\noindent where we introduced the parameter $A$ similar to that in
Ref. \cite{Sve03}, which is defined as

\begin{equation}\label{defA}
    A=\left(\frac{3\pi}{4}\frac{c}{v_F}\frac{\omega_p^2}
    {\omega_a^2}\tau\right)^{1/3}.
\end{equation}

\noindent In Eq. (\ref{Gdef}) near the lower integration limit
$y\sim\tau$ the reflection coefficient tends to 1 because the
argument of the function $F$ is in the Leontovich region ($b\gg1$)
where $(y/\sqrt{y^2-\tau^2})F(A/\sqrt{y^2-\tau^2}) \sim \tau/A\ll1$.
For this reason the contribution to the integral from the region
nearby the lower limit will be $\sim\tau^2$. One can neglect this
contribution changing the lower limit by zero and neglecting
$\tau^2$ in $\sqrt{y^2-\tau^2}$. Introducing the integration
variable $x=y/A$ one finds

\begin{equation}\label{Gs}
    G_s(\tau)=A^2 \int\limits_{0}^{\infty }dxx\ln\left(1-r_s^2(1/x)
    e^{-Ax}\right)+O(\tau^2),
\end{equation}

\noindent where

\begin{equation}\label{rs}
    r_s\simeq-\frac{1-F(1/x)}{1+F(1/x)}.
\end{equation}

This integral can be analyzed in two limit cases $A\ll1$ and
$A\gg1$. The former case is realized for extremely low temperatures
($T\ll 0.1^{\circ}K$ for $a>100\ nm$) and should be used to check
the behavior of the entropy at $T\rightarrow0$. At all realistic
temperatures the latter case is realized. Let us consider this case
first. At $A\gg1$ the main contribution to the integral Eq.(\ref{Gs})
comes from the region $x\sim1/A$. Then the argument of $F(1/x)$ is
large and we are in the Leontovich impedance region, where
$F(1/x)\simeq 4x/3\sqrt{3}$. This situation was already described in
Ref. \cite{Sve03} and the result can be written immediately:

\begin{equation}\label{GsAgg1}
    G_s(\tau)=-\zeta(3)+\frac{4}{3\sqrt{3}}
    \frac{8}{A}\zeta(3)+O(1/A^2),\ A\gg1.
\end{equation}

\noindent Here an additional factor $4/3\sqrt{3}$ in comparison with
Eq. (29) in Ref. \cite{Sve03} takes into account different
definitions of $A$ used in this paper. Two other terms in Eq.
(\ref{delF}) can be easy calculated with the help of Eq.
(\ref{GsAgg1}).

In the opposite limit $A\ll1$ the situation is different. In this
case, the important values of $x$ in the integral (\ref{Gs}) are $x\sim1$
and the Leontovich approximation is no longer valid. In this limit the
exponent can be changed by 1 and the integral in Eq. (\ref{Gs}) is
just a number that can be found numerically substituting expression
for $F(1/x)$ from Eq. (\ref{FGdef}) into Eq. (\ref{rs}). The result
will be the following:

\begin{equation}\label{GsAll1}
    G_s(\tau)=-0.0938A^2+O(A^3),\ A\ll1.
\end{equation}

\noindent The important difference of this expression from that
found in Ref. \cite{Sve03} (see Eq. (23) therein) is that
$G_s\rightarrow0$ when $A$ is going to zero instead of the finite
value $G_s\rightarrow-\zeta(3)$. The reason for this change of the
behavior is the reflection coefficient. When the Leontovich
approximation is used in Eq.(\ref{rs}) $r_s\rightarrow1$ at
$A\rightarrow0$ but the use of the exact impedance (\ref{Zs}) gives
$r_s\rightarrow0$.

Now let us find the function $G_p(\tau)$. As in the case of
$G_s(\tau)$ one can neglect $\tau^2$ in $\sqrt{y^2-\tau^2}$ and
change the lower integration limit by zero. In the low frequency
(temperature) range the first term in Eq. (\ref{Zp}) dominates and
the reflection coefficient can be presented as

\begin{equation}\label{reflp}
    r_p=\frac{1-\frac{1}{\sqrt{3}}\frac{v_F}{c}\frac{\omega_a}{\omega_p}y}
    {1+\frac{1}{\sqrt{3}}\frac{v_F}{c}\frac{\omega_a}{\omega_p}y}\approx
    1-\frac{2}{\sqrt{3}}\frac{v_F}{c}\frac{\omega_a}{\omega_p}y.
\end{equation}

\noindent Typically the reflection coefficient for $p$-polarization
is approaching 1 in the low frequency region. Small correction in
Eq. (\ref{reflp}) appears as a nonlocal effect connected with the
Thomas-Fermi screening. With this $r_p$ the integral in Eq.
(\ref{Gdef}) is easily calculated and for $G_p(\tau)$ one finds

\begin{equation}\label{Gp}
    G_p(\tau)=-\zeta(3)\left(1-\frac{8}{\sqrt{3}}\frac{v_F}{c}
    \frac{\omega_a}{\omega_p}\right).
\end{equation}

\noindent It does not depend on $\tau$ at all and holds true in both
limits of large and small $A$. The same conclusion was made in Ref.
\cite{Sve03} but without the Thomas-Fermi correction.

Now we are able to present the final expressions for the temperature
dependent part of the free energy in the limits $A\ll1$ and $A\gg1$.
Calculating the integrals in Eq. (\ref{delF}) using the functions
Eq.(\ref{GsAll1}), Eq.(\ref{Gp}) and the definition of the $n=0$ term in
Eq.(\ref{zeroterm}) one finds in the limit of small $A$:

\begin{equation}\label{delFAll1}
    \Delta {\cal F}\left( a,T\right) =\frac{kT}{ 8\pi a^{2}}\left[
    -\alpha \zeta(3)+\frac{1}{2}\zeta(3)\left( 1-\frac{8
    v_F\omega_a}{\sqrt{3}c \omega_p}\right) + 0.0146 A^2+O(A^3)
    \right],\ \ \ A\ll1,
\end{equation}

\noindent where the first term, containing $\alpha$, originates from
the $n=0$ term ${\cal F}_{0}\left( a,T\right)$. This expression is different
 from Eq. (28) in Ref. \cite{Sve03}, where the Leontovich
impedance was used. First, the coefficient 1/2(1 + Thomas-Fermi
correction) in front of the $\zeta$-function shows that only $p$
polarization contributes to the $A$-independent part and, second,
the $A$-dependent part behaves as $A^2$ instead of $A\ln A$. These
changes are due to different behavior of the exact impedance
Eq.(\ref{Zs}) in comparison with the Leotovich impedance (\ref{Leont}).
On the contrary, for $A\gg1$ the Leontovich impedance is a good
approximation and we successfully reproduce Eq. (33) of Ref.
\cite{Sve03}:

\begin{equation}\label{delFAgg1}
    \Delta {\cal F}\left( a,T\right) =\frac{kT}{ 8\pi a^{2}}
    \zeta(3)\left[-\alpha + 1-\frac{4
    v_F\omega_a}{\sqrt{3}c \omega_p}-\frac{32}{3\sqrt{3}}\left
    (\frac{1-2p_1}{A}\right)+O(A^{-2}) \right],\ \ \ A\gg1,
\end{equation}

\noindent where $p_1$ is a numerical coefficient the same as in Ref.
\cite{Sve03}, $p_1=0.0133$. The only new feature in this relation is
the presence of the Thomas-Fermi correction. Note that in the case
of large $A$ both polarization contribute equally to the $A$
independent term (1/2+1/2+ Thomas-Fermi correction).

\section{Entropy and Discussion \label{sec5}}
Before discussing the entropy behavior in the low temperature limit
we should fix the parameter $\alpha$ in Eq. (\ref{zeroterm}) for the
$n=0$ term. Let us separate it in two parts describing $s$ and $p$
polarizations

\begin{equation}\label{alpha}
    \alpha=\alpha_s+\alpha_p.
\end{equation}

\noindent Contribution of $p$ polarization in the classical part of
the free energy ${\cal F}_{0}\left( a,T\right)$ is not problematic.
As we know the only new feature that appeared due to nonlocality is
the Thomas-Fermi screening. It has clear physical meaning and should
present in any reasonable approach. To find $\alpha_p$ we has to
take the impedance (\ref{Zp}) at $\zeta \rightarrow 0$ and calculate
the function $-G_p(0)/2$ (see Eq. (\ref{Gdef})). But we already
found the function $G_p(\tau)$, which is given by Eq. (\ref{Gp}) and
in our approximation it does not depend on $\tau$ at all, therefore,

\begin{equation}\label{alpha_p}
    \alpha_p=\frac{1}{2}\left(1-\frac{8}{\sqrt{3}}\frac{v_F}{c}
    \frac{\omega_a}{\omega_p}\right).
\end{equation}

\noindent It is important that the Thomas-Fermi correction in
$\alpha_p$ is exactly canceled with that in the $A$-independent part
of the free energy (\ref{delFAll1}) or (\ref{delFAgg1}). Therefore,
the Thomas-Fermi screening finally does not contribute to the
temperature dependent part of the free energy.

The real problem is connected with the value of $\alpha_s$. In
Bostr\"{o}m and Sernelius approach \cite{Bos00} $s$-polarization
does not contribute to the $n=0$ term and $\alpha_s=0$. When SDM
prescription is used for the $n=0$ term \cite{Sve00} the
contribution of $s$ polarization is the same as for the ideal metal:
$\alpha_s=1/2$. The plasma model prescription for the $n=0$ term
\cite{Bor00} gives $\alpha_s=\alpha_s(\omega_a/\omega_p)$ as a
function of the separation, which approaching 1/2 at
$\omega_a/\omega_p\ll1 $. Close value of $\alpha_s$ gives
extrapolation of the Leontovich impedance from the infrared range to
zero frequency used in Ref. \cite{Gey03}. Different values of
$\alpha_s$ are responsible for different temperature corrections in
these approaches.

At very low temperature when $A\ll1$ the entropy calculated from Eq.
(\ref{delFAll1}) is

\begin{equation}\label{SAll1}
    S=-\frac{\partial \Delta {\cal F}}{\partial T}=\frac{k}
    {8\pi a^2}\left[\alpha_s \zeta (3)-\frac{5}{3}0.0146
    A^2\right],\ \ \ A\ll 1.
\end{equation}

\noindent It goes to zero at $T\rightarrow 0$ only if $\alpha_s=0$.
In this case the entropy approaches zero from the negative side as
$T^{2/3}$. This conclusion coincides with that made in Refs.
\cite{Hoy03,Bos04} on the basis of finite residual resistivity. The
use of finite $\omega_{\tau}(0)$ was criticized in Ref. \cite{Bez04}
(see also a recent preprint \cite{Bez05}) on the ground that the
Nernst heat theorem was formulated for equilibrium states and any
defects in the material responsible for the residual resistivity
should be considered as deviation from equilibrium. The objection is
reasonable but here we showed that the residual resistivity did not
play physical role at low temperatures. Instead the nonlocal effects
are responsible for the effective relaxation frequency (see Eq.
(\ref{Etanom})) $v_Fk\sim v_F/a\sim 10^{13}\ rad/s$, which is much
more important than tiny $\omega_{\tau}(0)$. Nevertheless, as Eq.
(\ref{SAll1}) demonstrates the final conclusion of Refs.
\cite{Hoy03,Bos04} holds true.

At higher temperatures when $A\gg1$ but $\tau \ll1$ the entropy is
still negative. In this range from Eq. (\ref{delFAgg1}) one finds
for the entropy

\begin{equation}\label{SAgg1}
    S=\frac{k}
    {8\pi a^2}\zeta (3)\left[\alpha_s - \frac{1}{2}+
    \frac{64}{9\sqrt{3}}\left(\frac{1-2p_1}{A}\right)\right],\
    \ \ A\gg 1,
\end{equation}

\noindent which is obviously negative for $\alpha_s=0$.

The entropy is a positively defined physical value and the
negative value for the Casimir entropy is puzzling. Recently
\cite{Bre04} some arguments were provided justifying the negative
Casimir entropy as long as the total entropy is positive.  The
free energy of the whole system consists of two contributions. The
main additive part comes from the short-range atomic interaction.
The long-range interaction realized via fluctuating fields gives
much smaller contribution to the free energy, but this
contribution can be separated due to its nonadditive character
(see discussion of this problem in Ref. \cite{LP9}). The additive
and nonadditive parts are independent on each other because the
first is defined by the volume but the second depends on the
separation between bodies. Usually it is assumed that the
nonadditive part is given by the Casimir free energy of
fluctuating fields. The idea proposed in Ref. \cite{Bre04} is that
part of the nonadditive free energy can belong to the bodies. In
this case one can write

\begin{equation}\label{nonad}
    \Delta {\cal F}\left( a,T\right)=\Delta {\cal F}_{body}\left(
    a,T\right)+\Delta {\cal F}_{field}\left( a,T\right)
\end{equation}

\noindent Both terms give contribution to the entropy

\begin{equation}\label{nonadentr}
    S\left(a,T\right)=-\frac{\partial\Delta {\cal F}_{body}}{\partial
    T}-\frac{\partial\Delta {\cal F}_{field}}{\partial T}
\end{equation}

\noindent The second term here is negative at low $T$ but the first
term could provide the total entropy to be positive. This idea can
be true but we would like to stress that the term $\Delta {\cal
F}_{body}\left(a,T\right)$ should be explicitly specified. This is
because it gives contribution not only to the entropy, but also to
the force according to the relation

\begin{equation}\label{nonadforce}
    F(a,T)=-\frac{\partial\Delta {\cal F}_{body}}{\partial
    a}-\frac{\partial\Delta {\cal F}_{field}}{\partial a}.
\end{equation}

\noindent At the moment we do not know any corrections to the
Casimir force which appear not from the fluctuating field but from
the nonadditive free energy of the bodies. In our opinion the
negative Casimir entropy is the evidence of a thermodynamic problem.
We should stress, however, that all the other approaches to the
temperature correction equally suffer the thermodynamic problem
because for $\alpha_s\neq0$ the entropy is finite at $T=0$.

On the other hand, the zero contribution of $s$ polarization to the
$n=0$ term has solid physical grounds. In the local case the
$1/\omega$ behavior of the dielectric function, responsible for
 the vanishing of the  reflection coefficient $r_s$, is the direct
result of the Ohm's law. Any attempts to change this behavior will
break this law. The plasma model describes well the infrared optics
but this is only an approximation, which cannot be used as a low
frequency limit as was proposed in Ref. \cite{Bor00}. Otherwise any
real metal would be a perfect conductor. The same is true for the
impedance approach, which is extrapolated from the infrared optics
to zero frequency \cite{Gey03,Bez04}, and for the SDM prescription
as in Ref. \cite{Sve00}. Our nonlocal analysis does not bring
anything new in the $n=0$ term because in the zero frequency limit
the impedance Eq.(\ref{Zs}) coincides with the exact local impedance.
It is known that properly defined local impedances reproduce the
force in the dielectric function approach \cite{Esq03}, therefore,
$\alpha_s=0$.

There is a very simple physical explanation why $s$ polarization
should not contribute to the force in the low frequency limit. If
$z$ is the normal direction to the metal surface, then $s$-polarized
field can be chosen as having the following nonzero components of
magnetic and electric fields: $H_x$, $H_z$, and $E_y$. When
$\omega\rightarrow0$ the magnetic field can be found from the
Maxwell equation $\nabla\times \textbf{H}=4\pi \textbf{j}/c$, where
$\textbf{j}$ is the external current density responsible for the
fluctuating fields \cite{LP9}. The electric field, which is
described by the equation $\nabla\times \textbf{E}=i\omega
\textbf{H}/c$, will be suppressed in comparison with $\textbf{H}$
because $\omega$ is small. So in the limit $\omega\rightarrow0$
$s$-polarized field degenerates to pure magnetic field. But the
magnetic field penetrates freely via nonmagnetic metals that means
that the reflection coefficient is going to zero.  Similarly the
$p$-polarized field degenerates to pure electric field in the
$\omega\rightarrow0$ limit. The electric field is screened by the
metal and the reflection coefficient is 1.

We came to a contradictory situation. From electrodynamics it
follows that $\alpha_s=0$. On the other hand, thermodynamics shows
that the Casimir entropy in this case is negative and something must
be wrong. All the other approaches proposed in the literature are
equally unsuccessful thermodynamically ($S\neq0$ at $T=0$) but, in
addition, they do not follow from electrodynamics. We cannot resolve
the thermodynamical problem by breaking the laws of electrodynamics.
Specifically we should stress that the approach based on
extrapolation of the Leontovich impedance from infrared to zero
frequency \cite{Bez04} cannot be accepted as physical. It disregards
$q$-dependence of the impedances, which plays crucial role for
evanescent field configurations. The authors postulated that the
evanescent fields have the same reflection coefficients as the
propagating fields. The Casimir effect is not the only physical
phenomenon where the evanescent fields can be probed. In the well
investigated domains like near field optics or near field microwaves
$q$ dependence plays principal role. No deviations from the standard
electrodynamics were noted so far.

To all appearance the experimental situation is not in favor
$\alpha_s=0$. This case contradicts to Lamoreaux experiment
\cite{Lam97}. Also there were claims that $\alpha_s=0$ does not
agree with the experiments by Decca et al. \cite{Dec03}. However,
very high roughness of metallic films in these experiments did not
allow these claims to be considered seriously. Recently \cite{Dec05}
the same group refined their measurements reducing the surface
roughness and increasing precision of determination of the absolute
separation. It is important that an experimental error of $0.6\%$
holds in a wide range of separations from 170 nm to 300 nm. However,
in this experiment no attempt was made to characterize the used gold
films optically. Instead, the handbook \cite{HB1} optical data were
used for calculation of the force. It was demonstrated that
\cite{Sve03b,Sve04b} that the optical data for gold films prepared
in different conditions can variate very significantly. Prediction
of the force with the precision better than 2\% should include
direct measurement of the optical properties of the films especially
in the mid-infrared range \cite{Sve04b}. Nevertheless, even with the
use of the handbook optical data one can conclude that the case
$\alpha_s=0$, probably, is not supported by the experiment. This is
because the handbook optical data present the best samples. The
unannealed films used in the experiment should have smaller
reflection coefficients than the handbook data predict. As the
result, the theoretical force was overestimated in Ref.
\cite{Dec05}. It means that the difference between the measured
force and predicted one in the case $\alpha_s=0$ can be only larger.
Of course, this is the result of only one group and one has to wait
an independent confirmation of it. It should be mentioned also that
the best way to see the temperature correction \cite{Bre04} is the
change of the temperature in the experiment.

All the discussion above shows that the situation with the thermal
correction to the Casimir force is in deep crisis. To the moment we
do not know any approach which is in agreement with both
electrodynamics and thermodynamics.

\section{Conclusions \label{Sec6}}
We analyzed behavior of the Casimir free energy at low temperatures.
The main contribution to the temperature dependent part of the free
energy $\Delta {\cal F}$ is defined by the low frequencies
$\zeta\sim2\pi kT/\hbar$ that is in contrast with the temperature
independent part, which is defined by the characteristic frequency
$\zeta\sim\omega_a$. With the temperature decrease the anomalous
skin effect becomes increasingly important for $\Delta {\cal F}$.
General theory of nonlocal impedances was used for calculations. It
was demonstrated that at low temperatures the relaxation frequency
does not play any physical role. Instead, the physical significance
get the frequency $v_F k$, where $k$ is the wave number. The
approximate Leontovich impedance describe the situation well if
$T\gg(v_F/c)(\hbar\omega_a/2\pi)$. When this condition is not satisfied one
cannot use the approximate Leontovich impedance any more.

The troubling $n=0$ term in the Lifshitz formula was parameterized
by the parameter $\alpha$ (see Eq. (\ref{zeroterm})), which is
different for different approaches to the temperature correction
discussed in the literature. This parameter was kept arbitrary in
calculations. In the temperature range $\hbar\omega_a(v_F/c)\ll 2\pi
kT\ll\hbar\omega_p(v_F/c)$ we reproduced for the free energy the
same result as in Ref. \cite{Sve03}, where the Leontovich impedance
of the anomalous skin effect was used. However, at smaller
temperatures, $2\pi kT\ll\hbar\omega_a(v_F/c)$, the behavior of
$\Delta {\cal F}$ drastically changes because dependence of the
impedance $Z_s$ on the transverse momentum $q$ becomes important. It
was demonstrated that the entropy is going to zero in the limit
$T\rightarrow0$ only in the case when $s$ polarization does not
contribute to the $n=0$ term ($\alpha_s=0$). In all other cases the
entropy is finite at $T=0$.

However, even in the case $\alpha_s=0$ the entropy at low
temperatures is negative that, in our opinion, indicates the
presence of the thermodynamic problem. It was demonstrated that the
idea on total positive entropy proposed in Ref. \cite{Bre04} at
least incomplete. We concluded that the thermal Casimir force is in
deep crisis and any approach to resolve the problem should respect
both the laws of thermodynamics and electrodynamics.

{\it Note added} After submission of this manuscript we became aware
of the work by Bo Sernelius, Phys. Rev. B (to be published) who also
analyzed the nonlocal effects in the Casimir problem. The conclusion
on the entropy behavior coincides with ours but the method of
analysis is different.

 \acknowledgements{We thank Rub\'{e}n Barrera for
helpful discussions. Partial support from CONACyT grant 44306 and DGAPA-UNAM grant IN101605.}


\begin{references}
\bibitem{Cas48}  H. B. G. Casimir, Proc. K. Ned. Akad. Wet. {\bf 51}, 793
(1948).

\bibitem{Lam97}  S. K. Lamoreaux, Phys. Rev. Lett. {\bf 78}, 5 (1997); {\bf %
81}, 5475 (1998).

\bibitem{Moh98}  U. Mohideen and A. Roy, Phys. Rev. Lett. {\bf 81}, 4549
(1998); A. Roy, C.-Y. Lin, and U. Mohideen, Phys. Rev. D {\bf 60},
111101(R) (1999); B. W. Harris, F. Chen, and U. Mohideen, Phys. Rev. A {\bf 62%
}, 052109 (2000).

\bibitem{Ede00}  T. Ederth, Phys. Rev. A 62, 062104 (2000).

\bibitem{Cha01}  H. B. Chan, V. A. Aksyuk, R. N. Kleiman, D. J. Bishop, and
F. Capasso, Science {\bf 291}, 1941 (2001); Phys. Rev. Lett. {\bf
87}, 211801 (2001).

\bibitem{Bre02}  G. Bressi, G. Carugno, R. Onofrio, and G. Ruoso, Phys.
Rev. Lett. {\bf 88}, 041804 (2002).

\bibitem{Dec03}  R. S. Decca, D. L\'{o}pez, E. Fischbach, and D. E. Krause,
Phys. Rev. Lett. {\bf 91}, 050402 (2003); R. S. Decca, E. Fischbach,
G. L. Klimchitskaya, D. E. Krause, D. L\'{o}pez, and V. M.
Mostepanenko, Phys. Rev. D {\bf 68}, 116003 (2003).

\bibitem{Dec05} S. Decca, D. L\'{o}pez, E. Fischbach, G. L. Klimchitskaya,
D. E. Krause,  and V. M. Mostepanenko, Ann. Phys., {\bf 318}, 37 (2005).

\bibitem{Ian04}  D. Iannuzzi, M. Lisanti, and F. Capasso, Proc. Natinonal
Acad. Sci. USA, {\bf 101}, 4019 (2004); M. Lissanti, D. Iannuzzi and
F. Capasso, arXiv:quant-ph/0502123 (2005).

\bibitem{Mil04} K. A. Milton, J. Phys. A: Math. Gen. \textbf{37},
R209 (2004).

\bibitem{Meh67} J. Mehra, Physica \textbf{37}, 145 (1967).

\bibitem{Bro69} L. S. Brown and G. J. Maclay, Phys. Rev.
\textbf{184}, 1272 (1969).

\bibitem{Lif56}  E. M. Lifshitz, Zh. Eksp. Teor. Fiz. {\bf 29}, 94 (1956)
[Sov. Phys. JETP {\bf 2}, 73 (1956)].

\bibitem{LP9}  E. M. Lifshitz and L. P. Pitaevskii, {\it Statistical
Physics, Part 2} (Pergamon Press, Oxford, 1980).

\bibitem{Sch78} J. Schwinger, L. L. DeRaad, and K. A. Milton, Ann.
Phys. (NY) \textbf{115}, 1 (1978).

\bibitem{Bos00} M. Bostr\"{o}m and B. E. Sernelius, Phys. Rev. Lett.
\textbf{84}, 4757 (2000).

\bibitem{Bor00} M. Bordag, B. Geyer, G. L. Klimchitskaya, and V. M.
Mostepanenko, Phys. Rev. Lett. \textbf{85}, 503 (2000).

\bibitem{Sve00} V. B. Svetovoy and M. V. Lokhanin, Mod. Phys. Lett.
A \textbf{15}, 1013 (2000); textbf{15}, 1437 (2000).

\bibitem{Sve01} V. B. Svetovoy and M. V. Lokhanin, Phys. Lett.
A \textbf{280}, 177 (2001).

\bibitem{Bez02a} V. B. Bezerra, G. L. Klimchitskaya, and V. M.
Mostepanenko, Phys. Rev. A \textbf{65}, 052113 (2002).

\bibitem{Sve03} V. B. Svetovoy and M. V. Lokhanin, Phys. Rev.
A \textbf{67}, 022113 (2003).

\bibitem{Hoy03} J. S. H{\o}ye, I. Brevik, J. B. Aarseth, and K. A.
Milton, Phys. Rev. E \textbf{67}, 056116 (2003).

\bibitem{Bos04} M. Bostr\"{o}m and B. E. Sernelius, Physica A
\textbf{339}, 53 (2004).

\bibitem{Bez02b} V. B. Bezerra, G. L. Klimchitskaya, and C. Romero,
Phys. Rev. A \textbf{65}, 012111 (2002).

\bibitem{Gey03} B. Geyer, G. L. Klimchitskaya, and V. M.
Mostepanenko, Phys. Rev. A \textbf{67}, 062102 (2003).

\bibitem{Bez04} V. B. Bezerra, G. L. Klimchitskaya, V. M.
Mostepanenko, and C. Romero, Phys. Rev. A \textbf{69}, 022119
(2004).

\bibitem{Sve04} V. B. Svetovoy, Phys. Rev. A \textbf{70}, 016101 (2004).

\bibitem{Gey04} B. Geyer, G. L. Klimchitskaya, and V. M.
Mostepanenko, Phys. Rev. A \textbf{70}, 016102 (2004).

\bibitem{Tor04} J. R. Torgerson and S. K. Lamoreaux, Phys. Rev. E
\textbf{70}, 047102 (2004).

\bibitem{Lam05} S. K. Lamoreaux, Rep. Prog. Phys. \textbf{68}, 201
(2005).

\bibitem{Moc02} W. L. Moch\'{a}n, C. Villarreal, and R. Esquivel-Sirvent,
Rev. Mex. Fis. \textbf{48}, 339 (2002).

\bibitem{Esq03} R. Esquivel-Sirvent, C. Villarreal, and W. L. Moch\'{a}n,
Phys. Rev. A \textbf{68}, 052103 (2003).

\bibitem{Esq05} R. Esquivel-Sirvent, C. Villarreal, and W. L. Moch\'{a}n, Phys. Rev. A {\bf 71}, 029904 (2005) 

\bibitem{Bre03} I. Brevik, J. Aarseth, J. S. H{\o}ye, and K. A. Milton,
in Quantum Field Theory Under the influence of External Conditions,
edited by K. A. Milton (Rinton Press, Princeton, 2004), p. 54,
arXiv: quant-ph/0311094.

\bibitem{Bre04} I. Brevik, J. Aarseth, J. S. H{\o}ye, and K. A.
Milton, arXiv: quant-ph/0410231.

\bibitem{Esq04} R. Esquivel, and V. B. Svetovoy,
Phys. Rev. A \textbf{69}, 062102 (2004).

\bibitem{Hein75} J. Heinrichs, Phys. Rev. B, {\bf 11} 3625 (1975). 

\bibitem{Kli68}  K. L. Kliewer and R. Fuchs, Phys. Rev. {\bf 172}, 607
(1968).

\bibitem{Abr88} A. A. Abrikosov, \emph{Fundamentals of the Theory of
Metals} (North-Holland, Amsterdam, 1988).

\bibitem{Bez05} V. B. Bezerra, R. S. Decca, E. Fischbach, B. Geyer,
G. L. Klimchtskaya, D. E. Krause, D. L\'{o}pez, V. M. Mostepanenko,
and C. Romero, arXiv: quant-ph/0503134.

\bibitem{HB1} \emph{Handbook of Optical Constants of Solids}, edited
by E. D. Palik (Academic Press, 1995).

\bibitem{Sve03b} V. B. Svetovoy, in Quantum Field Theory Under the influence of External Conditions,
edited by K. A. Milton (Rinton Press, Princeton, 2004), p. 76,
arXiv: cond-mat/0401562.

\bibitem{Sve04b} V. B. Svetovoy, arXiv: cond-mat/0412123.


\end{references}
\end{document}